\documentclass[twocolumn]{revtex4-1}
\usepackage{graphicx}
\usepackage{color}
\newcommand{\hH}{\hat{{\cal H}}}

\newcommand{\ket}[1]{\left| #1 \right\rangle}
\newcommand{\bra}[1]{\left\langle #1 \right|}
\newcommand{\braket}[2]{\left\langle #1 | #2 \right\rangle}

\newcommand{\proj}[1]{| #1\rangle\!\langle #1 |}

\newcommand{\eea}{\end{eqnarray}}
\newcommand{\bea}{\begin{eqnarray}}
\newcommand{\ee}{\end{equation}}
\newcommand{\be}{\begin{equation}}

\newcommand{\ZoZ}{Z\!\!\!Z}

\begin{document}
\title{Exchanging identical particles \& topological quantum computing}

\author{S.J. van Enk}

\affiliation{Department of Physics and
Oregon Center for Optical, Molecular \& Quantum Sciences\\
University of Oregon, Eugene, OR 97403}

\begin{abstract}
The phase factor $(-1)^{2s}$ that features in the exchange symmetry for identical spin-$s$ fermions or bosons is not simply and automatically equal to the phase factor one can observe in an interference experiment that involves physically exchanging two such particles. The observable phase contains, in general,  single-particle geometric and dynamical phases as well, induced by both spin and spatial exchange transformations. By extending the analysis to (non-abelian) anyons it is argued that, similarly, there are single-anyon geometric and dynamical contributions in addition to purely topological unitary transformations that accompany physical exchanges of anyons.
Work remains to be done in order to demonstrate---if it is still true---that those additional contributions to the gates in anyonic topological quantum computers do not destroy the inherent robustness of the ideal gates. 
This negative result is described most clearly in terms of the Berry matrix.
\end{abstract}

\maketitle

\section{Introduction}
Suppose we have a properly (anti-)symmetrized state of two identical fermions/bosons of the form
\bea\label{Psi}
\ket{\Psi}_{1,2}=(2)^{-1/2}
\left[\ket{\alpha}_1\ket{\beta}_2+(-1)^{2s}\ket{\beta}_1\ket{\alpha}_2\right]
\eea
where $\ket{\alpha}$ and $\ket{\beta}$ are orthogonal single-particle  states and $s$ denotes the spin of each particle. 
The ``exchange symmetry'' possessed by the state $\ket{\Psi}$ of Eq.~(\ref{Psi}) is expressed by the mathematical identity
\be\label{tt}
\ket{\Psi}_{2,1}=(-1)^{2s}\ket{\Psi}_{1,2}.
\ee
Many textbooks describe the content of the symmetry (\ref{tt}) by the statement that when the two particles are exchanged, their state acquires a phase factor $(-1)^{2s}$. Now it is true that exchanging the two Hilbert space labels $1\leftrightarrow 2$ yields the transformation $\ket{\Psi}_{1,2}\longmapsto(-1)^{2s}\ket{\Psi}_{1,2}$.
Exchanging labels, however, is a mathematical operation, not a physical one. It may be the case that physically exchanging the two particles (by actually moving them around)  yields a different transformation, even though the transformed state still must satisfy (\ref{tt}). 

Similarly, in the position representation, changing coordinates by swapping $\vec{r}_1$ and $\vec{r}_2$ (in configuration space \cite{leinaas1977}) of a wave function $\Psi(\vec{r}_1,\vec{r}_2)$
is a mathematical operation and physically swapping two particles does not necessarily simply result in the transformation
$\Psi(\vec{r}_1,\vec{r}_2)\longmapsto\Psi(\vec{r}_2,\vec{r}_1)$. The importance of distinguishing between the {\em mathematical} and the {\em physical} content of exchange symmetry was emphasized long ago in, e.g., Ref.~\cite{mirman1973}. This distinction is a recurring theme in the current paper.
As we will see, there is more to physically swapping particles than there is to swapping mathematical
labels or coordinates. 

The questions considered here are: What are the possible observable phase factors the quantum state (\ref{Psi}) could undergo if we apply a physical operation that takes state $\ket{\alpha}$ to $\ket{\beta}$ and vice versa (as opposed to switching labels 1 and 2)? How does such a phase depend on the physical implementation of this exchange? Discussion of the answers raises two further questions: What is the status of the exchange phase of abelian {\em anyons} \cite{wilczek1982}? And how does this discussion generalize to the much more complicated case of non-abelian anyons? The answers are of fundamental interest but are of special importance in the context of anyonic quantum computation \cite{kitaev2003,nayak2008,pachos2012}.

The above questions were inspired by a recent proposal
\cite{roos2017} to measure the exchange phase
for two identical ions (which could be fermions or bosons, depending on what isotope is chosen) trapped on opposite sides of a ring.
The ions can be physically swapped by rotating each ion over an angle $\pi$ (in the same direction) 
 around the center of the ring.  As in any interference experiment, what can be measured is a difference between two phases: one accumulated during the swapping process, the other accumulated when the particles stay in place. For convenience we may assume the latter phase to be equal to zero.  (The idea is that phases accrued while transporting the particles are in practice (much) harder to control than phases that accrue when the particles are merely idling.)

In the proposed experiment \cite{roos2017} the ions not only start off in orthogonal spatial states but remain so during the entire particle swapping process. That is, they are always in non-overlapping spatial regions, and so in principle (and in practice!) the two ions can be addressed individually (by using focused laser beams) at all times.
In this sense they remain distinguishable. This condition of distinguishability is always assumed in the context of quantum computing where we need the ability to address individual qubits and keep track of which one is which, even if those qubits are all identical particles, and even if those qubits are encoded in non-local degrees of freedom. (For different perspectives on the difference between the concepts ``identical'' and ``indistinguishable,'' see Refs.~\cite{de1975,dieks1,dieks2,van2018}.)

We consider the straightforward case of identical fermions and identical bosons in quite some detail in the next Section, mostly to motivate and prepare for  the discussion in Section \ref{anyons} of abelian and non-abelian anyons.

\section{Identical fermions and bosons}
\label{onetwo}
We turn to the questions posed in the Introduction about two identical particles in the state (\ref{Psi}).
For simplicity we specialize here to (single-particle) states of the form
\bea\label{AB}
\ket{\alpha}&=&\ket{A}\ket{\sigma},\nonumber\\
\ket{\beta}&=&\ket{B}\ket{\tau},
\eea
where for each particle we write the spatial state first and the spin state second.
For the spatial states we assume $\braket{A}{B}=0$, but  for the spin states their overlap $|\braket{\sigma}{\tau}|$ could be any  number between zero and unity (we ignore the trivial case $s=0$ here and assume $s\geq 1/2$).
Note that we can perfectly well say that the particle that is in the spatial state $\ket{A}$ has spin state $\ket{\sigma}$ but {\em not} that it is particle '1' that has spin state $\ket{\sigma}$.  We may thus label the physical particles by $j=A,B$, i.e., by their initial positions, but for this to work at later times we have to assume the particles will remain distinguishable  \footnote{Note here that ``classical particles'' emerge from quantum theory as approximately localized entities \cite{dieks2}, not as entities labeled '1' or '2'.}.

Let us focus on Hamiltonian (pure-state) evolution of the two-particle state.
In order to preserve the symmetry of the state (\ref{Psi}) the two-particle Hamiltonian $\hH_{1,2}$ must be symmetric under an exchange of the labels 1 and 2. For simplicity we consider the case of two identical particles that do not interact with each other (this is sufficient, though not necessary, for our purpose). In that case, the Hamiltonian must have the form
\be\label{12}
\hH_{1,2}=\hat{H}_1\otimes\openone_2+
\openone_1\otimes\hat{H}_2
\ee
with $\hat{H}$ an arbitrary single-particle Hamiltonian and $\openone$ the identity operator
on the single-particle Hilbert space.

 \subsection{Spatial swapping}
We first focus on the swapping of spatial states and ignore effects from swapping the spin states. That is, in this subsection we simply set $\ket{\sigma}=\ket{\tau}$, and we assume these spin states to not change during the spatial swapping process.

Consider a physical operation, starting at time $t=0$ and ending at time $t=T$, that swaps the particles'  spatial states such that the particles remain in orthogonal spatial states during the whole process.
One way to accomplish this is to use a Hamiltonian that has no spatial dependence (and so does not distinguish between particles $j=A,B$). For a simple example, let us model the above-mentioned two ions trapped in a ring by two particles moving around a circle in the $x,y$ plane whose position is indicated by the polar angle $\phi$ (i.e., just one spatial degree of freedom; this is, once again, sufficient for our purpose here). We could 
use a time-dependent Hamiltonian  of the form
\be\label{H12}
\hH_{1,2}(t)=\dot{\phi}(t) \hat{L}_z^{(1)}\otimes \openone_2 +(1\leftrightarrow 2),
\ee
with $\phi(t)$ an arbitrary differentiable function of time with $\phi(0)=0$ and $\phi(T)=\pi$,  and with $\hat{L}_z$ the operator for the single-particle orbital angular momentum w.r.t. the center of the circle. Since  the two particles do not interact with each other, it suffices to consider the single-particle evolution
\be
\ket{A}\mapsto
\ket{A(t)}:=
\sum_m \braket{m}{A}\exp(-im\phi(t))\ket{m},
\ee
in terms of the eigenstates $\ket{m}$ of $\hat{L}_z$
with $m\in \ZoZ$ (and similarly for $\ket{B(t)}$).
The requirement that the state $\ket{A}$
transform into $\ket{B}$ at time $t=T$ 
implies
\be
\ket{B}=\ket{A(T)}=
\sum_m \braket{m}{A}\exp(-im\pi))\ket{m}.
\ee 
As a consequence we have
\be
\ket{B(T)}=\sum_m \braket{m}{A}\exp(-2im\pi))\ket{m}=\ket{A}.
\ee
The requirement that $\ket{A(t)}$ be orthogonal (at all times) to $\ket{B(t)}$ is fulfilled when it is true at time $t=0$, i.e., when
\be
\sum_m |\braket{m}{A}|^2\exp(im\pi)=0.
\ee
The single-particle unitary transformation \footnote{Path ordering is not necessary here as Hamiltonians at different times commute with each other.}
\be
\hat{U}=\exp\left(-i\int_0^T\!\!dt \hat{H}(t)/\hbar\right)
=\exp(-i\pi \hat{L}_z/\hbar)
\ee
thus has the simple properties
\bea\label{phasAB}
\hat{U}\ket{A}&=&\ket{B},\nonumber\\
\hat{U}\ket{B}&=&\ket{A}.
\eea
This implies that we obtain the transformation 
\be\label{sss}
\ket{\Psi}_{1,2}\longmapsto \hat{U}_1\otimes\hat{U}_2\ket{\Psi}_{1,2}=(-1)^{2s}\ket{\Psi}_{1,2} 
\ee
for the two-particle state, seemingly confirming the idea that no matter how we physically exchange the two particles, we get a phase factor $(-1)^{2s}$.

 The reason for this apparent confirmation, however, is that a rotation of the coordinate system over an angle $\theta$ around the $z$ axis can be mathematically implemented by using the same unitary operator
$\exp(i\theta\hat{L}_z)$ we used here to represent the physical swapping operation. The {\em physical} particle swapping operation here thus happens to correspond to the {\em mathematical} operation of swapping coordinates or swapping labels.

However, we can add an extra term to the Hamiltonian (\ref{H12}) of the form 
\be
\hat{H}_{1,2}^{\rm extra}=\hbar\dot{\varphi}(t)\openone_1\otimes\openone_2
\ee
with $\varphi(t)$ an arbitrary differentiable function of time. We can, in addition, assume this Hamiltonian to be nonzero only during the swapping process, and, moreover, that it affects only the swapping process (i.e., it is absent when the particles idle \footnote{For example, one can make the spatial evolution dependent on the spin states of the particles, as in \cite{roos2017}. Section \ref{spin} spells out how to implement the converse, spin evolution conditioned on the spatial state.}).
The particles will still be physically swapped but the two-particle state will accumulate an additional dynamical phase equal to $\varphi_{{\rm spatial}}:=\varphi(0)-\varphi(T)$. That is, we find
instead of (\ref{sss}) the transformation
\be\label{spat}
\ket{\Psi}_{1,2}\longmapsto \exp\left[i\varphi_{{\rm spatial}}\right](-1)^{2s}\ket{\Psi}_{1,2}.
\ee
And so the presence of this dynamical phase (of which the authors of \cite{roos2017} are well aware) already implies that an actual measurable phase difference arising from physically exchanging two particles does not simply and automatically equal the exchange phase $(-1)^{2s}$.  

For future reference, we may also say that the family of operators
$\exp(i\theta\hat{L}_z)$ (for different values of $\theta$) form a unitary representation of the 1D rotation group in the 2D plane [a {\em mathematical} property], but that these operators are not the only unitary operators that implement a {\em physical} rotation.
There are more rigid restrictions on mathematical coordinate transformations than there are on physical operations.

\subsection{Spin swapping}\label{spin}
More interesting possibilities open up (namely, geometric phases) when we include the spin degrees of freedom.
In this subsection we assume $\braket{\sigma}{\tau}\neq 0$, and so in order to swap states $\ket{\alpha}\leftrightarrow\ket{\beta}$, we have to exchange the spin states of the particles as well.
We can still use the same Hamiltonian (\ref{H12}) for implementing the swap operation of the spatial degrees of freedom, and we just have to add a term that acts on the spin degrees of freedom to implement a spin swap.
We use the same generic form (\ref{12}) for the two-particle Hamiltonian,
where now $\hat{H}$ acts on both spatial and spin degrees of freedom of a single particle.
$\hat{H}$ can be different for a particle in state $\ket{A(t)}$ than it is for a particle in state $\ket{B(t)}$ (recall these are orthogonal states for all $t$).  For example, we can choose a single-particle time-dependent Hamiltonian of the form
\be
\hat{H}=\hat{\Pi}_{A}(t)\hat{H}_A(t)+\hat{\Pi}_{B}(t)
\hat{H}_B(t)\ee
with $\hat{H}_A(t)$ and $\hat{H}_B(t)$ acting exclusively on the spin degree of freedom, and with
projectors $\hat{\Pi}_{A,B}(t)$ defined as
\bea
\hat{\Pi}_A(t)&=&\proj{A(t)},\nonumber\\
\hat{\Pi}_B(t)&=&\proj{B(t)}.
\eea
We may thus introduce two different and independent (single-particle) unitary operators
$\hat{V}_A, \hat{V}_B$ acting on the spin degree of freedom,  which are generated by the different Hamiltonians $\hat{H}_A$ and $\hat{H}_B$, respectively, such that
\bea\label{phasess}
\hat{V}_A\ket{\sigma}&=&\exp(i\varphi_A)\ket{\tau},\nonumber\\
\hat{V}_B\ket{\tau}&=&\exp(i\varphi_B)\ket{\sigma},
\eea
with arbitrary phases $\varphi_{A,B}$.
Substituting  relations (\ref{AB}), (\ref{spat}) and (\ref{phasess}) into the  transformation of the  initial state (\ref{Psi})
yields 
\be\label{sum}
\ket{\Psi}_{1,2}\longmapsto
\exp(i(2s\pi+\varphi_{\rm spatial}+\varphi_{\rm spin}))\ket{\Psi}_{1,2}.
\ee
The total phase acquired in this process is written as a sum of three terms. Apart from the exchange phase $2s\pi$ and the spatial dynamical phase we mentioned in the previous subsection, we also have a phase
\be\label{spinphase}
\varphi_{\rm spin}=\varphi_A+\varphi_B,
\ee 
resulting from the combined effect of the two unitary spin transformations $\hat{V}_{A,B}$ 
\bea
\hat{V}_B\hat{V}_A\ket{\sigma}&=&\exp(i(\varphi_A+\varphi_B))\ket{\sigma}\nonumber\\
\hat{V}_A\hat{V}_B\ket{\tau}&=&\exp(i(\varphi_A+\varphi_B))\ket{\tau}.
\eea
This equation shows that the phase 
$\varphi_{\rm spin}$ can be interpreted as pertaining to single-particle trajectories around a closed loop in state space, the traditional setting for defining dynamical and geometric phases \cite{berry1984}. [The dynamical phase depends on how fast the system traverses the loop, the geometric phase depends only on the shape of the loop.]
The spin phase in general contains both types of phase, and its value depends on how the two unitary spin operations are implemented. One way is to choose 
$\hat{V}_A=\hat{V}_B^\dagger$ such that the total spin-dependent phase automatically vanishes, $\varphi_{\rm spin}=0$.
In that case, the total phase acquired in the particle-swapping process would equal the exchange phase if the spatial dynamical phase can be eliminated, and so the exchange phase $2s\pi$ would indeed be directly observable in an interference experiment, just as envisioned in Ref.~\cite{roos2017}.

On the other hand, if we would rotate both spins in the same way, i.e., such that $\hat{V}_A=\hat{V}_B$, then a full rotation of the spin around a fixed axis (applying to the case where $\ket{\sigma}$ would be spin ``up'' along a certain direction orthogonal to the rotation axis and $\ket{\tau}$  spin ``down'' along that same direction) would produce a geometric phase equal to $2s\pi$, thus equaling (and hence canceling) the exchange phase. In this case, there would be {\em no} observable difference between physically exchanging fermions or bosons!

The fact that the geometric phase for a full spin rotation equals the exchange phase, was used in Ref.~\cite{berry}  in an attempt to explain the spin-statistics theorem.
Here, though, we see that the two phases are conceptually different and that they both  contribute independently to an observable phase difference.

\subsection{Summary}
To summarize this Section, note the following property of the three terms appearing in the transformation (\ref{sum}). The spatial and spin phases are determined by the Hamiltonian evolution of the particles and each can take on {\em any} value, but the exchange phase $2s\pi$ is fixed for---or intrinsic to---pairs of identical spin-$s$ particles. 

The fundamental difference between the intrinsic exchange phase and the extrinsic geometric and dynamical phases can be best formulated as follows.
Consider  a quantum state of two identical spin-$s$ particles, $\ket{\Psi}$.
In the coordinate-spin representation this state 
may be written as
\be
\Psi(\vec{r}_1,\sigma_1;\vec{r}_2,\sigma_2):=
\braket{\vec{r}_1,\sigma_1;\vec{r}_2,\sigma_2}{\Psi}.
\ee
But the representation in terms of coordinates and spins labeled '1' and '2' is redundant when the particles are identical \cite{mirman1973,leinaas1977}. This redundancy in the representation is expressed as
\be
\bra{\vec{r}_1,\sigma_1;\vec{r}_2,\sigma_2}=
(-1)^s\bra{\vec{r}_2,\sigma_2;\vec{r}_1,\sigma_1},
\ee
and from this the usual expression of exchange symmetry (\ref{tt}) follows directly.
That is, the {\em same} state $\ket{\Psi}$ can be represented in two different but closely related (and, of course, equivalent) ways. If we change the representation once more by swapping labels again, we necessarily return to the same representation. 

In contrast, consider a change in the physical state by letting it evolve under some unitary evolution,
\be
\ket{\Psi}\longmapsto \hat{U}\ket{\Psi}.
\ee
This physical evolution gives rise to dynamical and geometric phase shifts. Physically swapping particles one way, and then swapping them back in another way, does not necessarily return the system to exactly the same state.

So, there is certainly a fundamental difference between the exchange phase and the geometric and dynamical phases. Because of the presence of the latter phases, the
measurable phase accompanying a physical exchange operation does not automatically equal the intrinsic exchange phase. 

\section{Anyons}\label{anyons}
A brilliant proposal for quantum computing is based on anyons \cite{kitaev2003,nayak2008}.
Anyons are quasi particles, emerging in certain 2D models embedded in our 3D world, describing electrons confined to move in a plane. Abelian anyons behave very much like particles with an intrinsic exchange phase {\em not} equal to an integer multiple of $\pi$. Non-abelian anyons undergo more complicated (non-commuting) unitary transformations when they are exchanged and this allows nontrivial operations on anyons to be performed by merely moving them around each other.
Let us look at this in more detail.
\subsection{Abelian anyons}
A simple model for an abelian anyon was proposed in \cite{wilczek1982}: a composite "particle" moving in the $x,y$ plane consisting of a magnetic flux $\Phi$ (with the magnetic field pointing in the $z$ direction) plus a charged particle, with charge $q$. This composite particle has the property that rotating it around its own axis gives rise to a nontrivial phase factor, equal to $\exp(iq\Phi/\hbar)$ due to the Aharonov-Bohm effect. Similarly, two such identical composite ``particles'' circling around each other exactly once both acquire half that phase, so that their joint two-particle state acquires the same phase $q\Phi/\hbar$. This is a {\em topological} phase (as opposed to geometric and dynamical phases) as it does not depend on how (by which path)  the anyons encircle each other (once), nor on how fast they do so. 

The Aharonov-Bohm phase, of course, is not the only phase acquired by a charged particle: in fact that phase is simply an add-on to the dynamical evolution the particle would undergo if the vector potential $\vec{A}$ were zero everywhere.
What is measured in an Aharonov-Bohm experiment is, indeed, a shift between two interference patterns, one observed when $\vec{A}=0$, the second observed in a fixed nonzero $\vec{A}$ field.

Consider now the following ``error'': some small area in the $x,y$ plane (let's call it $S$) with a small nonzero external magnetic field.
Now the phase a single anyon acquires  depends
on how many times its path goes around $S$. That dependence is still purely topological in nature. However, the phase also depends on whether the anyon passes through $S$.
If it does, the total magnetic flux through the area enclosed by the anyon's path  includes a finite fraction of the magnetic flux through $S$. This introduces a {\em very} weak geometric (non-topological) dependence.
Since that geometric addition to the total  phase is acquired only for a very small fraction of paths---only those that cut through $S$--- 
we may to an extremely good approximation consider the total phase acquired by an anyon to be purely topological in character.
That is, for almost all paths, the total phase is invariant under small continuous deformations of the path. In other words, the Aharonov-Bohm effect is robust against any local perturbation.

Next consider the case where there is a small magnetic field (which may be spatially varying) {\em everywhere}, i.e., in the whole $x,y$ plane.
The phase an anyon acquires consists of two parts now, one purely topological having to do with how many times it encircled other anyons, the second part  is purely geometric, depending on the external magnetic flux through the area enclosed by its path. Since the latter geometric contribution is no longer necessarily small, the Aharonov-Bohm effect is not robust against local perturbations everywhere. 

The topological phase the simple model anyons acquire when we move them around each other (exactly once)
is very similar in character to the exchange phase $2\pi s$ intrinsic to identical fermions/bosons, in the sense that its value is fixed by the property of the anyons themselves (the magnetic flux and the charge they carry).
However, just as for identical fermions/bosons, physically exchanging two anyons in a realistic situation produces an observable phase shift that includes, in general, nonzero dynamical and geometric phases. 
\subsection{Non-abelian anyons}
When two non-abelian anyons are swapped their joint state acquires more than just a phase shift. 
In fact, if there is a $D$-dimensional subspace of degenerate states (with the same or very nearly the same energy) available to a set of anyons, a unitary transformation $\hat{U}\in U(D)$
accompanies the physical exchange of two (or more) such anyons.
Topological quantum computing relies on the inherent robustness of the topological part of $\hat{U}$. The crucial question is, are there additional nontrivial (that is, not just an overall phase factor) non-topological contributions to
this $\hat{U}$ that are, therefore, not inherently robust?
There are two ways of describing the transformations of anyonic states: one in terms of unitary representations of the braiding group---a description inherently topological, but not complete---the other in terms of a higher-dimensional (non-abelian) generalization of the Berry phase. 
\subsubsection{Unitary representations of the braid group}
The most elegant explanation, topological in nature, of how anyonic quantum computing works, is in terms of the braid group. The motion of 2 particles in a 2D plane in time can be represented by braids: 2 lines connecting two points in the plane at the initial time to two points in the plane at the final time. The state space of identical anyons carries a unitary representation of the braid group.
That is, one can assign a unitary transformation
$\hat{B}_k\in U(D)$  to each element $B_k$ of the braid group such that $\hat{B}_l\hat{B}_m=\hat{B}_k$ whenever $B_lB_m=B_k$.
Moreover, as it turns out, one can choose a set $\{\hat{B}_k\}$ of such unitary transformations that forms a set of gate operations that is universal for quantum computation. Different braids {\em only} differ in topological properties, and hence this set $\{\hat{B}_k\}$ is purely topological in nature.

However,  all this does not imply that the unitary state transformation $\hat{V}_k$ that results from a particular way of physically braiding two anyons, with the braid represented mathematically by $B_k$, necessarily equals the corresponding unitary $\hat{B}_k$.
There are three reasons (not quite independent) for this: (1) the operator
$\hat{B}_k$ describes a mathematical transformation, not the complete physical transformation (as we will see in more detail in the next subsection and echoing the conclusion arrived at in Section \ref{onetwo}),
(2) quantum mechanical evolution allows for an additional phase factor, (3) the unitary representation is not unique. 

Concerning the third reason, note the following.
Given unitary operators
$\{\hat{B}_k\}$, the operators  
$\{\hat{V}^\dagger\hat{B}_k\hat{V}\}$ for a fixed unitary operator $\hat{V}$ form a unitary representation of the braiding group, too.
Now, if we would actually implement
the transformation
$\hat{V}^\dagger\hat{B}_k\hat{V}$ with the same $\hat{V}$ appearing for all possible braiding operations, then this would merely amount to a trivial basis change. 
However, since the anyons must be distinguishable (in the context of quantum computing) and since the Hamiltonian of the actual system may vary over time,  it may well be that for a given physical braiding operation involving a particular pair of anyons $C$ and $D$ at some time $t$, $\hat{V}$ depends on that pair $C,D$ and on $t$. In particular, it may depend on the paths along which those two anyons are exchanged at that particular time.
This possibility of nontrivial path dependence has to be checked in an actual calculation. Let us see how such a calculation indeed leaves this possibility open.

\subsubsection{The Berry matrix}
We sketch here the calculation of Ref.~\cite{bonderson2011} concerning anyons arising in the fractional quantum Hall effect (electrons moving in a 2D plane in the presence of a strong perpendicular magnetic field with magnitude $B$. See also Refs.~\cite{prodan2009,wu2014}). 
Consider (excited) states 
parametrized by the positions $\vec{\lambda}_j$  (in the $x,y$ plane) of $2n$ quasi-particles for $j=1\ldots 2n$.
In particular, we may define states $\{\ket{\phi_a(\vec{\lambda}_j)}\}$ for $a=1\ldots D$ that form an orthonormal basis of the D-dimensional degenerate subspace of eigenstates  of the Hamiltonian  $H(\vec{\lambda}_j)$ for one particular fixed energy. (Here, $D=2^n-1$ \cite{nayak}.) That is, for each different value of $a$ we have a {\em different} state with all quasi particles at the same positions.  The state (but not its energy) depends on in what way two groups of $n$ quasi particles are associated with the coordinates of one pair of electrons \footnote{The wave functions are anti-symmetrized over all electron coordinates, but not over the positions of the quasi-particles, which are merely parameters.}. 

One way to describe topological quantum computing based on non-abelian anyons is by calculating explicitly the adiabatic evolution
a system undergoes by slowly changing the  control parameters $\vec{\lambda}_j$.
One crucial part of the calculation involves the Berry connection (or the Berry matrix) \cite{berry1984}
\be
({\cal A}_i)_{ab}=i \bra{\phi_a(\vec{\lambda}_j)}\frac{\partial}{\partial \vec{\lambda}_j}\ket{\phi_b(\vec{\lambda}_j)}.
\ee
We integrate the connection over a closed loop $L$ in parameter space (the parameter values $\vec{\lambda}_j$ all return to their initial values) to get a unitary evolution operator
\be\label{loopL}
\hat{U}_L={\cal P}\exp\left[
i\sum_i\oint_L {\cal A}_i(\vec{\lambda}_j) \cdot d\vec{\lambda}_j\right],
\ee
where the ${\cal P}$ symbol stands for path ordering, necessary because ${\cal A}(\vec{\lambda}_j)$
and ${\cal A}(\vec{\lambda}'_i)$ do not necessarily commute with each other.  As long as the evolution is indeed adiabatic (an assumption granted here), $\hat{U}_L$ does not depend on how fast the loop in parameter space is traversed.

In Ref.\cite{bonderson2011} a nice set of $D$ wavefunctions is found that is (almost, up to corrections that are exponentially small in the distances between the quasi particles) orthonormal.
The unitary operator $\hat{U}_L$ does depend on the choices of initial and final basis states. By explicitly including the unitary basis transformation $\hat{B}$ between those two bases, one finds that the product $\hat{U}_L\hat{B}$ is gauge (basis-choice) independent. 
One very convenient choice of bases for explaining and calculating the physical effects of braiding the paths of two anyons around each other, and one that also 
allows us here to see clearly where undesirable non-topological contributions may arise, is one where $\hat{B}$ is determined purely by the mathematical properties of the wave function transformations induced by swapping coordinates (the ``monodromy'' transformation). The latter transformations form a unitary representation of the braid group and do not depend on either dynamical or geometric features of any path. With this convenient choice of basis wave functions, it is only the operator $\hat{U}_L$ that depends on the path $L$, but it does so in a very simple and innocuous way: $\hat{U}_L$ is just the overall Aharonov-Bohm phase factor multiplied by the identity matrix within the $D$-dimensional degenerate subspace. 
(None of this is trivial \cite{bonderson2011}.)

This path independence is remarkable---given the explicit path dependence of the definition (\ref{loopL})---and relies on a special feature of the underlying theory.
Namely, the wave functions mentioned above depend only on a particular combination of the positions $\vec{\lambda}_j$, namely
on
$\eta_j=(x_j+iy_j)/l_B$ but not on $\eta_j^*=(x_j-iy_j)/l_B$ with $x_j$ and $y_j$ here denoting the $x$ and $y$ coordinates of the position $\vec{\lambda}_j$ and with $l_B=\sqrt{\hbar/eB}$ the magnetic length.
In this way, the calculation of the Berry matrix (up to the Aharonov-Bohm phase) corresponding to the motion of one quasi particle moving around the other quasi particles
can be reduced to a line integral along a closed loop in the complex plane of a holomorphic function. Such an integral depends only on poles of the holomorphic function enclosed by that loop. Hence the purely topological nature of $\hat{U}_L\hat{B}$.

The reason for this very nice property is that the base theory starts with an idealized situation of a perfectly homogeneous magnetic field in the direction perpendicular to the $x,y$ plane, and non-interacting electrons confined to move in that plane.
Landau's description of that situation \cite{tong} is most elegantly given in terms of a bosonic annihilation operator
$\hat{a}$ that contains a derivative w.r.t. the coordinate $z^*=x-iy$. Ground states are then described by wave functions $\Psi$ such that $\hat{a}\Psi=0$. That equation gives rise to infinitely many solutions \footnote{This is thanks to the high symmetry of the problem arising from the translational and rotational invariance of the infinite plane.}: any holomorphic function $f(z)$ of $z=x+iy$ multiplied by a Gaussian  factor $\exp(-|z|^2/(4l_B^2))$ works. 
It's the latter (non-holomorphic) Gaussian factor that, when integrated along a closed loop in the complex plane, gives rise to the only path dependence in the total unitary transformation, the Aharonov-Bohm phase.

Now, however, consider ``errors,'' i.e., deviations from the idealized system.
There are two features of an actual physical system that may ruin the topological robustness of the physical operator $\hat{U}_L\hat{B}$, features which, as far as the author knows, have not been considered previously in this context (non-adiabaticity has been considered in Ref.~\cite{knapp2018} for example, but that is not the issue here.).

First, a base Hamiltonian for the electrons that differs {\em everywhere} from the ideal Hamiltonian
 \footnote{Of course, the presence of such global errors, as is well known, affects conventional quantum computers based on, say, ions in an ion trap or superconducting qubits just as well. Both types of qubits are localized (unlike anyonic qubits) and errors are typically described, therefore, as local errors. But in this case, too, since {\em all} qubits are locally affected by errors, the total Hamiltonian differs from the ideal Hamiltonian {\em everywhere}.} may not leave the wave functions purely holomorphic (aside from the usual Gaussian factor). For example, if the magnetic field is almost homogeneous, but not quite, and it has small nonzero $x$ and $y$ components, and if there are small stray electric fields everywhere, then there will be small corrections to the eigen-energy wave functions that may well be non-holomorphic.
For example, a function of $(x+iy)/l_B$ is no longer holomorphic if $l_B$ depends nontrivially on $x$ and $y$. 
 
 Second, possibly of greater concern, the interaction that allows one to move quasi-particles must constitute a strong deviation from the base Hamiltonian (and that is well-known, of course). What has been assumed in calculations like those in  Refs.~\cite{prodan2009,wu2014} is that the additional time-dependent Hamiltonian that drives the motion of the quasi-particles depends {\em only} on  the parameters $\vec{\lambda}_j$, and in fact, {\em only} on the combination $\zeta=x_j+iy_j$.
But errors in that time-dependent Hamiltonian may well depend on one additional parameter (which could be $\zeta^*$, but it could be any other parameter, say, a local laser intensity, a local magnetic field, a local electric field, etc., as well).
 That one extra parameter (which we may assume to vanish both at the beginning of the adiabatic process and at the end) changes the loop over which we integrate in (\ref{loopL}) in a drastic way. Namely, by changing the dimension of the parameter space, the topology of the class of closed loops is changed, too.  (Indeed, anyons exist only in 2D, not in 3D; for the Berry matrix, the additional parameter doesn't have to be the third spatial coordinate in order for the path dependence of (\ref{loopL}) to become nontrivial.)
 
Proofs of topological robustness would have to be extended to include these two types of deviations from the ideal Hamiltonian \footnote{
We focused here on the anyons appearing in the fractional quantum Hall effect.
Refs.~\cite{cheng2011,zyuzin2013,pedro2015,knapp2018} point out dephasing errors in a different model for topological quantum computing, one based on Majorana braiding.}
\footnote{One may wonder whether experiments  such as Ref.~\cite{pan2018} that claim to have demonstrated that anyonic braiding 
statistics is topologically robust, refute the point made here? The answer is no: that experiment only compares two phases resulting from two particular {\em discrete} ways  of braiding. The two ways of braiding are physically implemented by applying single-qubit $X$ operations to two different sets of qubits, one consisting of 4 qubits, the other of 6 qubits. This procedure does not demonstrate anything about topological robustness.}.

\section{Discussion and conclusions}

When two particles are physically exchanged, their joint state undergoes a unitary transformation that consists of three parts: a dynamical phase, which depends on how fast the particles are exchanged, a geometric part, which depends only on geometric features of the paths taken by the two particles (e.g., the area or the magnetic flux enclosed by the paths), and a topological part depending only on topological features of the paths (e.g., how many times the particles wind around each other).
The latter part is, by definition, invariant under continuous deformations of the paths, and, therefore, tends to be quantized (e.g., taking on the values $\pm 1$ for identical fermions or bosons moving in 3 spatial dimensions).

Quantum computing with anyons (moving in 2 spatial dimensions) is based on a beautiful idea, that the topological part of the unitary transformation can act as a gate; and that different topological transformations can form a set of gates that is universal for quantum computation.

There are two reasons for suspecting the unitary transformation induced by physically braiding anyons not to be topologically robust, even if the ideal transformation is. First, when the actual background Hamiltonian differs from the ideal Hamiltonian {\em everywhere} in space, the integrand appearing in the Berry matrix becomes a function of more variables than just the intended one, and the Berry matrix will thus, in general, become path dependent. Second, if the time-dependent Hamiltonian that causes the anyons to move around each other depends on more variables than just the positions of the moving anyons, then, again, the Berry matrix becomes, in general, a path-dependent quantity.

Thanks to Michael Raymer for useful comments on  various versions.

\bibliography{bosons1}

\begin{thebibliography}{31}%
\makeatletter
\providecommand \@ifxundefined [1]{%
 \@ifx{#1\undefined}
}%
\providecommand \@ifnum [1]{%
 \ifnum #1\expandafter \@firstoftwo
 \else \expandafter \@secondoftwo
 \fi
}%
\providecommand \@ifx [1]{%
 \ifx #1\expandafter \@firstoftwo
 \else \expandafter \@secondoftwo
 \fi
}%
\providecommand \natexlab [1]{#1}%
\providecommand \enquote  [1]{``#1''}%
\providecommand \bibnamefont  [1]{#1}%
\providecommand \bibfnamefont [1]{#1}%
\providecommand \citenamefont [1]{#1}%
\providecommand \href@noop [0]{\@secondoftwo}%
\providecommand \href [0]{\begingroup \@sanitize@url \@href}%
\providecommand \@href[1]{\@@startlink{#1}\@@href}%
\providecommand \@@href[1]{\endgroup#1\@@endlink}%
\providecommand \@sanitize@url [0]{\catcode `\\12\catcode `\$12\catcode
  `\&12\catcode `\#12\catcode `\^12\catcode `\_12\catcode `\%12\relax}%
\providecommand \@@startlink[1]{}%
\providecommand \@@endlink[0]{}%
\providecommand \url  [0]{\begingroup\@sanitize@url \@url }%
\providecommand \@url [1]{\endgroup\@href {#1}{\urlprefix }}%
\providecommand \urlprefix  [0]{URL }%
\providecommand \Eprint [0]{\href }%
\providecommand \doibase [0]{http://dx.doi.org/}%
\providecommand \selectlanguage [0]{\@gobble}%
\providecommand \bibinfo  [0]{\@secondoftwo}%
\providecommand \bibfield  [0]{\@secondoftwo}%
\providecommand \translation [1]{[#1]}%
\providecommand \BibitemOpen [0]{}%
\providecommand \bibitemStop [0]{}%
\providecommand \bibitemNoStop [0]{.\EOS\space}%
\providecommand \EOS [0]{\spacefactor3000\relax}%
\providecommand \BibitemShut  [1]{\csname bibitem#1\endcsname}%
\let\auto@bib@innerbib\@empty
\bibitem [{\citenamefont {Leinaas}\ and\ \citenamefont
  {Myrheim}(1977)}]{leinaas1977}%
  \BibitemOpen
  \bibfield  {author} {\bibinfo {author} {\bibfnamefont {J.~M.}\ \bibnamefont
  {Leinaas}}\ and\ \bibinfo {author} {\bibfnamefont {J.}~\bibnamefont
  {Myrheim}},\ }\href@noop {} {\bibfield  {journal} {\bibinfo  {journal} {Il
  Nuovo Cimento B (1971-1996)}\ }\textbf {\bibinfo {volume} {37}},\ \bibinfo
  {pages} {1} (\bibinfo {year} {1977})}\BibitemShut {NoStop}%
\bibitem [{\citenamefont {Mirman}(1973)}]{mirman1973}%
  \BibitemOpen
  \bibfield  {author} {\bibinfo {author} {\bibfnamefont {R.}~\bibnamefont
  {Mirman}},\ }\href@noop {} {\bibfield  {journal} {\bibinfo  {journal} {Il
  Nuovo Cimento B (1971-1996)}\ }\textbf {\bibinfo {volume} {18}},\ \bibinfo
  {pages} {110} (\bibinfo {year} {1973})}\BibitemShut {NoStop}%
\bibitem [{\citenamefont {Wilczek}(1982)}]{wilczek1982}%
  \BibitemOpen
  \bibfield  {author} {\bibinfo {author} {\bibfnamefont {F.}~\bibnamefont
  {Wilczek}},\ }\href@noop {} {\bibfield  {journal} {\bibinfo  {journal} {Phys.
  Rev. Lett.}\ }\textbf {\bibinfo {volume} {49}},\ \bibinfo {pages} {957}
  (\bibinfo {year} {1982})}\BibitemShut {NoStop}%
\bibitem [{\citenamefont {Kitaev}(2003)}]{kitaev2003}%
  \BibitemOpen
  \bibfield  {author} {\bibinfo {author} {\bibfnamefont {A.~Y.}\ \bibnamefont
  {Kitaev}},\ }\href@noop {} {\bibfield  {journal} {\bibinfo  {journal} {Ann.
  Phys.}\ }\textbf {\bibinfo {volume} {303}},\ \bibinfo {pages} {2} (\bibinfo
  {year} {2003})}\BibitemShut {NoStop}%
\bibitem [{\citenamefont {Nayak}\ \emph {et~al.}(2008)\citenamefont {Nayak},
  \citenamefont {Simon}, \citenamefont {Stern}, \citenamefont {Freedman},\ and\
  \citenamefont {Sarma}}]{nayak2008}%
  \BibitemOpen
  \bibfield  {author} {\bibinfo {author} {\bibfnamefont {C.}~\bibnamefont
  {Nayak}}, \bibinfo {author} {\bibfnamefont {S.~H.}\ \bibnamefont {Simon}},
  \bibinfo {author} {\bibfnamefont {A.}~\bibnamefont {Stern}}, \bibinfo
  {author} {\bibfnamefont {M.}~\bibnamefont {Freedman}}, \ and\ \bibinfo
  {author} {\bibfnamefont {S.~D.}\ \bibnamefont {Sarma}},\ }\href@noop {}
  {\bibfield  {journal} {\bibinfo  {journal} {Rev. Mod. Phys.}\ }\textbf
  {\bibinfo {volume} {80}},\ \bibinfo {pages} {1083} (\bibinfo {year}
  {2008})}\BibitemShut {NoStop}%
\bibitem [{\citenamefont {Pachos}(2012)}]{pachos2012}%
  \BibitemOpen
  \bibfield  {author} {\bibinfo {author} {\bibfnamefont {J.~K.}\ \bibnamefont
  {Pachos}},\ }\href@noop {} {\emph {\bibinfo {title} {Introduction to
  topological quantum computation}}}\ (\bibinfo  {publisher} {Cambridge
  University Press},\ \bibinfo {year} {2012})\BibitemShut {NoStop}%
\bibitem [{\citenamefont {Roos}\ \emph {et~al.}(2017)\citenamefont {Roos},
  \citenamefont {Alberti}, \citenamefont {Meschede}, \citenamefont {Hauke},\
  and\ \citenamefont {H{\"a}ffner}}]{roos2017}%
  \BibitemOpen
  \bibfield  {author} {\bibinfo {author} {\bibfnamefont {C.}~\bibnamefont
  {Roos}}, \bibinfo {author} {\bibfnamefont {A.}~\bibnamefont {Alberti}},
  \bibinfo {author} {\bibfnamefont {D.}~\bibnamefont {Meschede}}, \bibinfo
  {author} {\bibfnamefont {P.}~\bibnamefont {Hauke}}, \ and\ \bibinfo {author}
  {\bibfnamefont {H.}~\bibnamefont {H{\"a}ffner}},\ }\href@noop {} {\bibfield
  {journal} {\bibinfo  {journal} {Phys. Rev. Lett.}\ }\textbf {\bibinfo
  {volume} {119}},\ \bibinfo {pages} {160401} (\bibinfo {year}
  {2017})}\BibitemShut {NoStop}%
\bibitem [{\citenamefont {De~Muynck}(1975)}]{de1975}%
  \BibitemOpen
  \bibfield  {author} {\bibinfo {author} {\bibfnamefont {W.}~\bibnamefont
  {De~Muynck}},\ }\href@noop {} {\bibfield  {journal} {\bibinfo  {journal}
  {Int. J. Theor. Phys.}\ }\textbf {\bibinfo {volume} {14}},\ \bibinfo {pages}
  {327} (\bibinfo {year} {1975})}\BibitemShut {NoStop}%
\bibitem [{\citenamefont {Dieks}(1990)}]{dieks1}%
  \BibitemOpen
  \bibfield  {author} {\bibinfo {author} {\bibfnamefont {D.}~\bibnamefont
  {Dieks}},\ }\href@noop {} {\bibfield  {journal} {\bibinfo  {journal}
  {Synthese}\ }\textbf {\bibinfo {volume} {82}},\ \bibinfo {pages} {127}
  (\bibinfo {year} {1990})}\BibitemShut {NoStop}%
\bibitem [{\citenamefont {Dieks}\ and\ \citenamefont
  {Lubberdink}(2011)}]{dieks2}%
  \BibitemOpen
  \bibfield  {author} {\bibinfo {author} {\bibfnamefont {D.}~\bibnamefont
  {Dieks}}\ and\ \bibinfo {author} {\bibfnamefont {A.}~\bibnamefont
  {Lubberdink}},\ }\href@noop {} {\bibfield  {journal} {\bibinfo  {journal}
  {Found. Phys.}\ }\textbf {\bibinfo {volume} {41}},\ \bibinfo {pages} {1051}
  (\bibinfo {year} {2011})}\BibitemShut {NoStop}%
\bibitem [{\citenamefont {van Enk}(2018)}]{van2018}%
  \BibitemOpen
  \bibfield  {author} {\bibinfo {author} {\bibfnamefont {S.~J.}\ \bibnamefont
  {van Enk}},\ }\href@noop {} {\bibfield  {journal} {\bibinfo  {journal} {arXiv
  preprint arXiv:1810.05147}\ } (\bibinfo {year} {2018})}\BibitemShut {NoStop}%
\bibitem [{Note1()}]{Note1}%
  \BibitemOpen
  \bibinfo {note} {Note here that ``classical particles'' emerge from quantum
  theory as approximately localized entities \cite {dieks2}, not as entities
  labeled '1' or '2'.}\BibitemShut {Stop}%
\bibitem [{Note2()}]{Note2}%
  \BibitemOpen
  \bibinfo {note} {Path ordering is not necessary here as Hamiltonians at
  different times commute with each other.}\BibitemShut {Stop}%
\bibitem [{Note3()}]{Note3}%
  \BibitemOpen
  \bibinfo {note} {For example, one can make the spatial evolution dependent on
  the spin states of the particles, as in \cite {roos2017}. Section \ref {spin}
  spells out how to implement the converse, spin evolution conditioned on the
  spatial state.}\BibitemShut {Stop}%
\bibitem [{\citenamefont {Berry}(1984)}]{berry1984}%
  \BibitemOpen
  \bibfield  {author} {\bibinfo {author} {\bibfnamefont {M.~V.}\ \bibnamefont
  {Berry}},\ }\href@noop {} {\bibfield  {journal} {\bibinfo  {journal} {Proc.
  R. Soc. Lond. A}\ }\textbf {\bibinfo {volume} {392}},\ \bibinfo {pages} {45}
  (\bibinfo {year} {1984})}\BibitemShut {NoStop}%
\bibitem [{\citenamefont {Berry}\ and\ \citenamefont {Robbins}(1997)}]{berry}%
  \BibitemOpen
  \bibfield  {author} {\bibinfo {author} {\bibfnamefont {M.~V.}\ \bibnamefont
  {Berry}}\ and\ \bibinfo {author} {\bibfnamefont {J.~M.}\ \bibnamefont
  {Robbins}},\ }\href@noop {} {\bibfield  {journal} {\bibinfo  {journal} {Proc.
  R. Soc. Lond. A}\ }\textbf {\bibinfo {volume} {453}},\ \bibinfo {pages}
  {1771} (\bibinfo {year} {1997})}\BibitemShut {NoStop}%
\bibitem [{\citenamefont {Bonderson}\ \emph {et~al.}(2011)\citenamefont
  {Bonderson}, \citenamefont {Gurarie},\ and\ \citenamefont
  {Nayak}}]{bonderson2011}%
  \BibitemOpen
  \bibfield  {author} {\bibinfo {author} {\bibfnamefont {P.}~\bibnamefont
  {Bonderson}}, \bibinfo {author} {\bibfnamefont {V.}~\bibnamefont {Gurarie}},
  \ and\ \bibinfo {author} {\bibfnamefont {C.}~\bibnamefont {Nayak}},\
  }\href@noop {} {\bibfield  {journal} {\bibinfo  {journal} {Phys. Rev. B}\
  }\textbf {\bibinfo {volume} {83}},\ \bibinfo {pages} {075303} (\bibinfo
  {year} {2011})}\BibitemShut {NoStop}%
\bibitem [{\citenamefont {Prodan}\ and\ \citenamefont
  {Haldane}(2009)}]{prodan2009}%
  \BibitemOpen
  \bibfield  {author} {\bibinfo {author} {\bibfnamefont {E.}~\bibnamefont
  {Prodan}}\ and\ \bibinfo {author} {\bibfnamefont {F.~D.~M.}\ \bibnamefont
  {Haldane}},\ }\href@noop {} {\bibfield  {journal} {\bibinfo  {journal} {Phys.
  Rev. B}\ }\textbf {\bibinfo {volume} {80}},\ \bibinfo {pages} {115121}
  (\bibinfo {year} {2009})}\BibitemShut {NoStop}%
\bibitem [{\citenamefont {Wu}\ \emph {et~al.}(2014)\citenamefont {Wu},
  \citenamefont {Estienne}, \citenamefont {Regnault},\ and\ \citenamefont
  {Bernevig}}]{wu2014}%
  \BibitemOpen
  \bibfield  {author} {\bibinfo {author} {\bibfnamefont {Y.-L.}\ \bibnamefont
  {Wu}}, \bibinfo {author} {\bibfnamefont {B.}~\bibnamefont {Estienne}},
  \bibinfo {author} {\bibfnamefont {N.}~\bibnamefont {Regnault}}, \ and\
  \bibinfo {author} {\bibfnamefont {B.~A.}\ \bibnamefont {Bernevig}},\
  }\href@noop {} {\bibfield  {journal} {\bibinfo  {journal} {Phys. Rev. Lett.}\
  }\textbf {\bibinfo {volume} {113}},\ \bibinfo {pages} {116801} (\bibinfo
  {year} {2014})}\BibitemShut {NoStop}%
\bibitem [{\citenamefont {Nayak}\ and\ \citenamefont {Wilczek}(1996)}]{nayak}%
  \BibitemOpen
  \bibfield  {author} {\bibinfo {author} {\bibfnamefont {C.}~\bibnamefont
  {Nayak}}\ and\ \bibinfo {author} {\bibfnamefont {F.}~\bibnamefont
  {Wilczek}},\ }\href@noop {} {\bibfield  {journal} {\bibinfo  {journal}
  {Nuclear Physics B}\ }\textbf {\bibinfo {volume} {479}},\ \bibinfo {pages}
  {529} (\bibinfo {year} {1996})}\BibitemShut {NoStop}%
\bibitem [{Note4()}]{Note4}%
  \BibitemOpen
  \bibinfo {note} {The wave functions are anti-symmetrized over all electron
  coordinates, but not over the positions of the quasi-particles, which are
  merely parameters.}\BibitemShut {Stop}%
\bibitem [{\citenamefont {Tong}(2016)}]{tong}%
  \BibitemOpen
  \bibfield  {author} {\bibinfo {author} {\bibfnamefont {D.}~\bibnamefont
  {Tong}},\ }\href@noop {} {\bibfield  {journal} {\bibinfo  {journal} {arXiv
  preprint arXiv:1606.06687}\ } (\bibinfo {year} {2016})}\BibitemShut {NoStop}%
\bibitem [{Note5()}]{Note5}%
  \BibitemOpen
  \bibinfo {note} {This is thanks to the high symmetry of the problem arising
  from the translational and rotational invariance of the infinite
  plane.}\BibitemShut {Stop}%
\bibitem [{\citenamefont {Knapp}\ \emph {et~al.}(2018)\citenamefont {Knapp},
  \citenamefont {Karzig}, \citenamefont {Lutchyn},\ and\ \citenamefont
  {Nayak}}]{knapp2018}%
  \BibitemOpen
  \bibfield  {author} {\bibinfo {author} {\bibfnamefont {C.}~\bibnamefont
  {Knapp}}, \bibinfo {author} {\bibfnamefont {T.}~\bibnamefont {Karzig}},
  \bibinfo {author} {\bibfnamefont {R.~M.}\ \bibnamefont {Lutchyn}}, \ and\
  \bibinfo {author} {\bibfnamefont {C.}~\bibnamefont {Nayak}},\ }\href@noop {}
  {\bibfield  {journal} {\bibinfo  {journal} {Phys. Rev. B}\ }\textbf {\bibinfo
  {volume} {97}},\ \bibinfo {pages} {125404} (\bibinfo {year}
  {2018})}\BibitemShut {NoStop}%
\bibitem [{Note6()}]{Note6}%
  \BibitemOpen
  \bibinfo {note} {Of course, the presence of such global errors, as is well
  known, affects conventional quantum computers based on, say, ions in an ion
  trap or superconducting qubits just as well. Both types of qubits are
  localized (unlike anyonic qubits) and errors are typically described,
  therefore, as local errors. But in this case, too, since {\protect \em all}
  qubits are locally affected by errors, the total Hamiltonian differs from the
  ideal Hamiltonian {\protect \em everywhere}.}\BibitemShut {Stop}%
\bibitem [{Note7()}]{Note7}%
  \BibitemOpen
  \bibinfo {note} {We focused here on the anyons appearing in the fractional
  quantum Hall effect. Refs.~\cite {cheng2011,zyuzin2013,pedro2015,knapp2018}
  point out dephasing errors in a different model for topological quantum
  computing, one based on Majorana braiding.}\BibitemShut {Stop}%
\bibitem [{Note8()}]{Note8}%
  \BibitemOpen
  \bibinfo {note} {One may wonder whether experiments such as Ref.~\cite
  {pan2018} that claim to have demonstrated that anyonic braiding statistics is
  topologically robust, refute the point made here? The answer is no: that
  experiment only compares two phases resulting from two particular {\protect
  \em discrete} ways of braiding. The two ways of braiding are physically
  implemented by applying single-qubit $X$ operations to two different sets of
  qubits, one consisting of 4 qubits, the other of 6 qubits. This procedure
  does not demonstrate anything about topological robustness.}\BibitemShut
  {Stop}%
\bibitem [{\citenamefont {Cheng}\ \emph {et~al.}(2011)\citenamefont {Cheng},
  \citenamefont {Galitski},\ and\ \citenamefont {Sarma}}]{cheng2011}%
  \BibitemOpen
  \bibfield  {author} {\bibinfo {author} {\bibfnamefont {M.}~\bibnamefont
  {Cheng}}, \bibinfo {author} {\bibfnamefont {V.}~\bibnamefont {Galitski}}, \
  and\ \bibinfo {author} {\bibfnamefont {S.~D.}\ \bibnamefont {Sarma}},\
  }\href@noop {} {\bibfield  {journal} {\bibinfo  {journal} {Phys. Rev. B}\
  }\textbf {\bibinfo {volume} {84}},\ \bibinfo {pages} {104529} (\bibinfo
  {year} {2011})}\BibitemShut {NoStop}%
\bibitem [{\citenamefont {Zyuzin}\ \emph {et~al.}(2013)\citenamefont {Zyuzin},
  \citenamefont {Rainis}, \citenamefont {Klinovaja},\ and\ \citenamefont
  {Loss}}]{zyuzin2013}%
  \BibitemOpen
  \bibfield  {author} {\bibinfo {author} {\bibfnamefont {A.}~\bibnamefont
  {Zyuzin}}, \bibinfo {author} {\bibfnamefont {D.}~\bibnamefont {Rainis}},
  \bibinfo {author} {\bibfnamefont {J.}~\bibnamefont {Klinovaja}}, \ and\
  \bibinfo {author} {\bibfnamefont {D.}~\bibnamefont {Loss}},\ }\href@noop {}
  {\bibfield  {journal} {\bibinfo  {journal} {Phys. Rev. Lett.}\ }\textbf
  {\bibinfo {volume} {111}},\ \bibinfo {pages} {056802} (\bibinfo {year}
  {2013})}\BibitemShut {NoStop}%
\bibitem [{\citenamefont {Pedrocchi}\ and\ \citenamefont
  {DiVincenzo}(2015)}]{pedro2015}%
  \BibitemOpen
  \bibfield  {author} {\bibinfo {author} {\bibfnamefont {F.~L.}\ \bibnamefont
  {Pedrocchi}}\ and\ \bibinfo {author} {\bibfnamefont {D.~P.}\ \bibnamefont
  {DiVincenzo}},\ }\href@noop {} {\bibfield  {journal} {\bibinfo  {journal}
  {Phys. Rev. Lett.}\ }\textbf {\bibinfo {volume} {115}},\ \bibinfo {pages}
  {120402} (\bibinfo {year} {2015})}\BibitemShut {NoStop}%
\bibitem [{\citenamefont {Song}\ \emph {et~al.}(2018)\citenamefont {Song},
  \citenamefont {Xu}, \citenamefont {Zhang}, \citenamefont {Wang},
  \citenamefont {Guo}, \citenamefont {Liu}, \citenamefont {Xu}, \citenamefont
  {Deng}, \citenamefont {Huang}, \citenamefont {Zheng} \emph
  {et~al.}}]{pan2018}%
  \BibitemOpen
  \bibfield  {author} {\bibinfo {author} {\bibfnamefont {C.}~\bibnamefont
  {Song}}, \bibinfo {author} {\bibfnamefont {D.}~\bibnamefont {Xu}}, \bibinfo
  {author} {\bibfnamefont {P.}~\bibnamefont {Zhang}}, \bibinfo {author}
  {\bibfnamefont {J.}~\bibnamefont {Wang}}, \bibinfo {author} {\bibfnamefont
  {Q.}~\bibnamefont {Guo}}, \bibinfo {author} {\bibfnamefont {W.}~\bibnamefont
  {Liu}}, \bibinfo {author} {\bibfnamefont {K.}~\bibnamefont {Xu}}, \bibinfo
  {author} {\bibfnamefont {H.}~\bibnamefont {Deng}}, \bibinfo {author}
  {\bibfnamefont {K.}~\bibnamefont {Huang}}, \bibinfo {author} {\bibfnamefont
  {D.}~\bibnamefont {Zheng}},  \emph {et~al.},\ }\href@noop {} {\bibfield
  {journal} {\bibinfo  {journal} {Phys. Rev. Lett.}\ }\textbf {\bibinfo
  {volume} {121}},\ \bibinfo {pages} {030502} (\bibinfo {year}
  {2018})}\BibitemShut {NoStop}%
\end{thebibliography}%

\end{document}